\documentclass[12pt,a4paper]{article}
\usepackage{graphicx}

\begin{document}
\advance \textwidth by 0.5cm
\parindent = 0pt
\setlength{\parskip}{0.5\baselineskip}

\begin{center}
    {\Large\bf
Computer simulation of inhibition-dependent binding in a neural network}
\bigskip

    {\large\bf
A.K.Vidybida}\medskip

{\it Bogolyubov Institute for Theoretical Physics\\
     Metrologichna str., 14-B\\
Kyiv 03680, Ukraine\\
E-mail: vidybida@bitp.kiev.ua}\bigskip

\end{center}

\def\be{\begin{equation}}
\def\ee{\end{equation}}
\def\P#1{{\bf P_#1}}


\bigskip
\bigskip

\hrule\medskip
{\small\parskip=1pt\parindent=0pt
{\bf Abstract.}
Reverberating dynamics of neural network is modeled on PC in order to
illustrate possible role of inhibition as binding controller in the network.
The network is composed of binding neurons.
 In the binding neuron model (Vidybida, 1998) the
degree of temporal coherence between synaptic inputs is decisive for
triggering, and slow inhibition is expressed in terms of the degree, which is
necessary for triggering. Two learning mechanisms are implemented in the
 network, namely, adjusting synaptic strength and/or propagation delays. By
means of forced playing of external pattern the network is taught to support
dynamics with disconnected and bound patterns of activity. By choosing either
high, or low inhibition one can switch between the disconnected and bound
patterns, respectively. This is interpreted as inhibition-controlled binding in
the network.}\medskip

\medskip

\hrule\medskip
\bigskip

\section{Introduction}

Information about the external world reaches brain being distributed in space
and time. It is as well distributed
over the sensory pathways of different
modalities.  Moreover, different features of a compact in space and time signal
of definite modality can be represented by means of activities in separate
neuronal populations (Giard et al., 1995).  At the same time, during perception
separate portions of information are some way brought together to represent
coherent objects. Mechanisms which ensure that
separate peaces of
information represented as disjointed in space and time neuronal activities
produce impression of single coherent object are known as feature linking, or
binding mechanisms (Llin\'as et al., 1994; von der Malsburg, 1999).  Similar situation is with storage
of conceptual knowledge and its retrieval (Damasio, 1989). In this case it is
proposed (Tranel, Damasio, Damasio, 1997).
 that the retrieval starts in high-order association
cortices and sub-cortical nuclei as activation of multiple spatially segregated
sites in which the knowledge pertinent to a particular concept is stored in
a non-explicit form. These activities in turn evoke activities in early sensory
cortices as well as in motor structures giving explicit images of a concept
being retrieved.

Thus, the recruitment of some intermediary regions due to
time-locked activities in another regions seems to be essential for binding
 (Damasio et al., 1996).
This process may happen at different spatial scales. For conceptual knowledge
the scale is comparable with the dimension of brain. For perception of
elementary components of visual image it could be the size of a few columns in
the visual cortex (Eckhorn et al., 1988). The spatially
smallest version of binding has
been proposed for interpretation of information processing in a single neuron
(Vidybida, 1998). In this case (Fig.\ref{schemec}) the spatial scale is
delimited by the set of synaptic inputs into a single cell.
If a number of inputs comes temporally coherent\footnote{For a fixed number,
$n$, of incoming into a neuron spikes, their
degree of temporal coherence, $TC$, is
defined in (Vidybida, 1998) as $TC=1/(t_n-t_1)$, where $t_1$, $t_n$ are the
arrival times of the first and the last spike, respectively.}, the cell fires a
spike, which is interpreted as binding of temporally coherent inputs into a
single event (the output spike).
  For a single cell it is proposed that slow inhibition could be a
factor that effectively controls this type of binding (Vidybida, 1998).

In this paper, the neuronal model of Fig.\ref{schemec} is utilized for
constructing a simple neuronal network. The reverberating dynamics of this
network is studied by means of computer simulation. The dynamical pattern in
which some geometrically intermediate region is recruited into firing due to
activity in adjacent regions is interpreted as a model for binding in this
network, because after the recruitment, the domain of activity becomes
spatially connected. The purpose of this paper is to demonstrate that slow
inhibition could control this type of binding in the network under
consideration.

\section{Methods}

The network composed of cells described in Fig.\ref{schemec} has been modeled
by means of C++ programming language and run on PC with time-step $dt=0.1$ ms.
Here are the details of the model used.

\subsection{Single cell}

\begin{figure}
\input{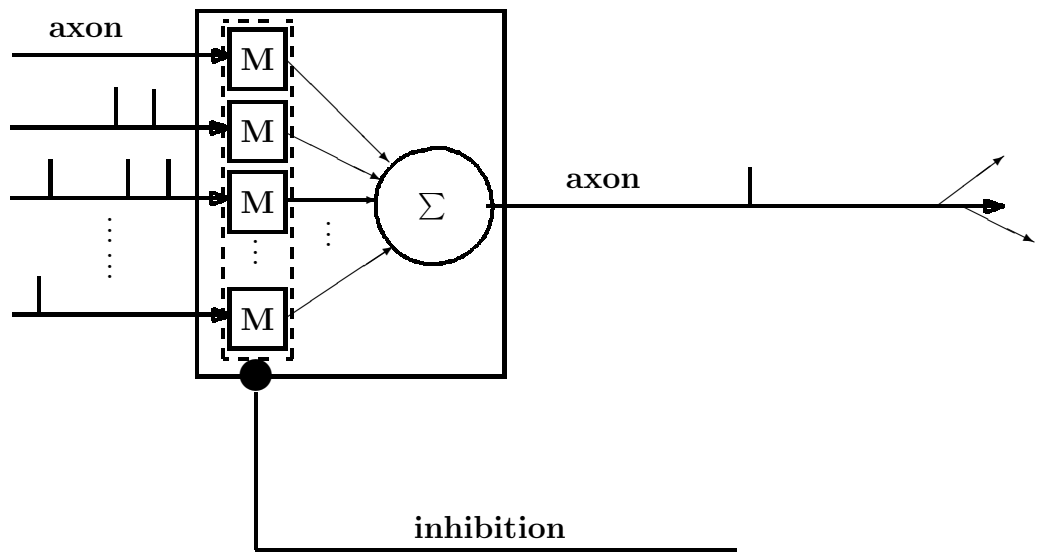}
\protect\caption{\label{schemec} The binding neuron.
The memory block {\bf M} for each input, mimics the EPSP finite lifetime.
It stores input spike during period of time $W$, and then forgets it.
 If the total
number of input
signals taken with their weights exceeds the
threshold value $BT$, the cell fires a spike.  Slow inhibition is expressed
in terms of the $W$: the higher is the inhibition, the narrower is the $W$. The
total number of inputs into a single cell depends on the connection pattern in
the network.} \end{figure}

Single neuron model is based on the previously made numerical simulation of
Hodgkin-Huxley-type neuron stimulated from multiple synaptic inputs
(Vidybida, 1996). This model 
is realized as a unit with a number of inputs for receiving
signals from other neurons, and an output for sending spikes to other neurons
(Fig.\ref{schemec}). The exact number of inputs depends on connection pattern
and varies from 4 for the case of the
nearest neighbor connections to $N-1$ for fully connected
network of $N$ neurons. Any received input
signal is stored (memorized) in the cell
during period $W$ after which it is forgotten. The temporal memorization
mimics the temporal behavior of the excitatory post-synaptic potential
(EPSP).
The width $W$ can be adjusted
for any input. Reduction of $W$ can be interpreted as increasing of
inhibition (Vidybida,1998).
Here the $W$ is chosen from the range 1-4 ms.
The cell is realized in such a way that the $W$ may be different for different
cells, and for different inputs in a single cell, but in
this work the $W$ is changed uniformly for all
inputs in all cells. Each input has its own weight (synaptic strength),
which can be changed during learning (see below).  The summation block ($\sum$
in Fig.\ref{schemec}) evaluates
at any moment of time the total amount of excitation stored in the
cell, which is the sum of all memorized inputs 
signals taken with their weights. If the
excitation reaches the binding (firing) threshold ($BT$), the cell fires a spike
and then stays in refraction for 0.9 ms.  Each cell has additional external
input the strength of which
just exceeds $BT$ (not shown in Fig.\ref{schemec}). As a
result a single spike arriving through the external input triggers
the cell immediately.  This enables a possibility to play external patterns
of activity which are used for teaching (learning).

\subsection{Axon}

Each axon is realized as a program unit characterized by the two cells which are
connected by this axon. Also it has its own length, which is equal to the
distance between the cells it connects. Axon receives signal from its input
cell, propagates it with velocity $v$ and eventually pumps it into its
output cell through the cell's synaptic input. At any moment of time, axon
either propagates a spike, or is waiting for a spike from its input cell.

\subsection{Network, general}

The spatial structure of network is fixed as a plane structure. For
constructing the network in a computer memory, the total number of neurons, $N$
should be specified. All $N$ units are arranged in a square lattice of
rectangular shape. The number of units in a row and in a column of the
rectangular are proportional to given numbers $H$ and $V$, respectively.
Depending on the numbers $N,H,V$, the last row in the square lattice
can be left incomplete.

In order to specify exact physical parameters of inter-neuronal connections,
the propagation speed in an axon, $v$ is fixed. The propagation delay between
any two nearest neighbor units, $d_p$ is fixed as well. This allows to
calculate the lattice constant $a$ (the distance between two nearest neighbor
units) as $a=vd_p$. With the exact value of $a$, one can determine
geometrical distance between any two units.

Axonal connection pattern is imposed as follows. A maximal admitted delay, $D$
is specified. If $D$ is equal or exceeds the propagation time along the
diagonal of the rectangular which represents the network,
 then any two cells are connected with two
identical axons propagating in opposite directions. Otherwise, only axons with
delays which are less or equal to $D$ are retained. For example, if $D=d_p$,
then only the nearest neighbors will be connected.
After construction, it is possible to specify
additionally which connections in the net should be
excitatory, and which are inhibitory.

During runtime, the program allows to pause the network dynamics and to make
some manipulations with its structural and dynamical properties and then
run the dynamics further. Namely, it is possible to choose two subsets of
neurons and to lock or unlock all axons, which connect any neuron from one
subset with any other neuron from the another subset. Axons propagating from
one subset to
the another one and those propagating in the opposite direction can
be manipulated separately.  Also it allows to play compulsory external patterns
of activity.  Patterns should be described by specifying which
groups of cells have to fire at each moment of time during the presence of
external drive, e.g. as in the Table 1.

\subsection{Network, realization}\label{NR}

A network of $15\times 15$ binding neurons has been used for modeling
reverberating dynamics.
The propagation speed, $v$ was set to 10 m/s.
Propagation delay between the nearest neighbors is
equal to 1 ms. This results in a square-shaped network with the square edge
length equal to 140 mm.  Connections with delay less or equal
to 4 ms are retained
in the network ($D=4$ ms). This brings about the connectivity pattern in which
a cell can be connected by reciprocal connections with up to 48 neighbors,
provided its location is far enough from the square boundary. The binding
threshold, $BT=300$. The synaptic strength is initially set to 10.

\begin{figure}
\input{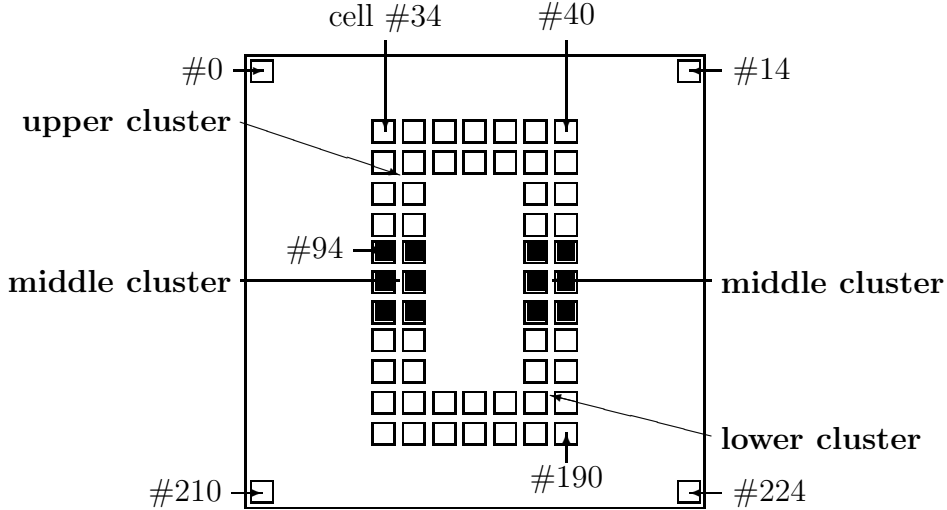}
\protect\caption{\label{o-net}
Domains in the network,  which display spatially bound/unbound activity
depending on the level of inhibition.
Neurons shown in black belong to the
middle clusters. These clusters start firing for low
inhibition and become silent for the inhibition made high. Cells
outside of the O-shaped domain (only 4 of them are displayed)
 remain silent during all stages of simulation.}
\end{figure}

Connections from the O-shaped region, which is shown in Fig.\ref{o-net}, to the
remaining part of the network are chosen inhibitory. This prevents activity in
the O-shaped region from spreading over the remaining part of the network. As
this is the only purpose of inhibition in this pull of connections, its
concrete realization does not matter in this work.  All connections inside of
the O-shaped region are excitatory.  In its initial state the network is at
rest.  This means that all axons are free of spikes, all cells are free of
input signals, and not in their refractory states. The pattern used to drive
the network (see Table \ref{pattern}) triggers only neurons in the O-shaped
region. In this case the firing activity is unable to spread outside of the
O-shaped region.  Thus, for dynamical states
which are studied here, the remaining part of the network may be considered as
cut from the O-shaped region, even if for another
tested driving patterns (not shown
here) intensive reciprocal dynamical interaction between the both parts may
happen.

\subsection{Learning}

Two types of learning are implemented in the network.
Both can be switched on and off during the runtime.
The first one is the Hebbian learning for
adjusting of synaptic strength. If the synaptic learning mode is switched on,
then every time the cell is triggered, the synaptic strength is advanced by 1
for those synapses which have contributed to this triggering.

The second one is adjusting the propagation (synaptic and/or axonal) delays. If
this mode is switched on, then if the cell receives a synaptic input during its
refractory period then the delay in corresponding input is reduced by $dt$.
A possible
biological counterpart of this learning rule is discussed in (Gerstner {\it at
al.}, 1996).

Both learning modes are useful for teaching the network to play a
certain dynamical pattern. For this purpose a text file should be prepared in
which the order in which the cells must fire is specified.
This file can be fed
into the network during the runtime. Each cell in this case receives
input through the additional external input which triggers cell
independently of
 contribution from other inputs. After sessions of teaching, the network
is able to play dynamics which may resemble the pattern used for teaching (see
n.\ref{Results}).

It appeared in preliminary experiments that
a network of this type with uniform patterns of synaptic strengths and
axonal connections, all of which are excitatory, is not suitable to play
prolonged dynamics in which some parts of network are silent. Instead, the
activity either spreads over the entire network, or dies out within tens of ms.
But, to discuss binding, one needs to have in the network both
spatially bound and
unbound patterns of activity, which are stable in a sense.  One
possibility is to make inhibitory all connections from one part to another
one (see n.\ref{NR}).  Another possibility is to teach the network to play
dynamics with activity in a limited domain.  This is possible by means of
playing a compulsory pattern in a domain of the network with
connections to the remaining part being locked,
 and with learning switched on. This allows
to modify connections in the domain in such a way that it becomes able to
support dynamics which in a sense is not very suitable for spreading further
after unlocking the locked axons. Both approaches were
used here in combination.

\section{Results}\label{Results}

The network was taught to support the dynamical pattern of sustained activity
in which 12 neurons in the middle clusters of the O-shaped region
(Fig.\ref{o-net}) are silent. For this purpose axons
propagating spikes from the upper and lower
clusters to the middle ones were locked during the teaching sessions. The
network was forced to play an external pattern which consists in serial firing
of groups of neurons in the upper and lower clusters (see Table \ref{pattern}).
\begin{table}
\begin{tabular}{ccccccccccccc}
\hline
time, ms& 0& 0.2& 0.4& 0.6& 0.8& 1.0& 1.2& 1.4& 1.6& 1.8& 2.0& 2.1\\
\hline
cell \#
&79  &64  &34  &49  &36  &37  &38  &40  &39  &69  &84  &-\\
&  80& 65 &50  &35  &51  &52  &53  &54  &55  &70  &85  &-\\
& 144& 159& 174& 175& 173& 172& 171& 170& 169& 154& 139&-\\
& 145& 160& 190& 189& 188& 187& 186& 184& 185& 155& 140&-\\
\hline
\end{tabular}
\caption{\label{pattern}Firing pattern used for teaching.
Four cells in each table column are triggered at the moment
specified above this column.
When the table
is exhausted, it is fed to the network again until the teaching time
is exhausted.
The cells are numbered as shown in Fig.\protect\ref{o-net}.
} \end{table}
In the first teaching session the
synaptic weights were modified, in the second one additionally the propagation
delays were modified.  Sessions duration was 100 ms and 200 ms. In the third
session the network was run freely for 100 ms with both learning mechanisms
enabled.  During the sessions, slow inhibition was low, which is expressed in
the value $W=4$ ms for each cell.  Initial state for any next session was the
final state of the previous one.

\unitlength 1mm
\begin{figure}
\begin{picture}(160,130)(0,20)

\put(10,120){\includegraphics[width=0.69\textwidth]{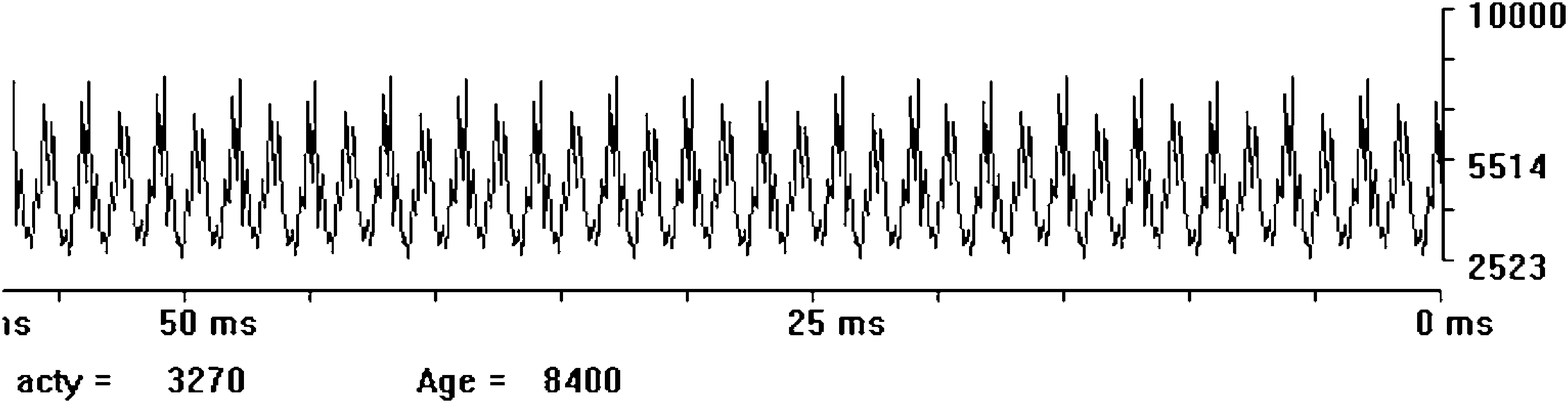}}
\put(0,115){\Large\it a}
\put(10,80){\includegraphics[width=0.69\textwidth]{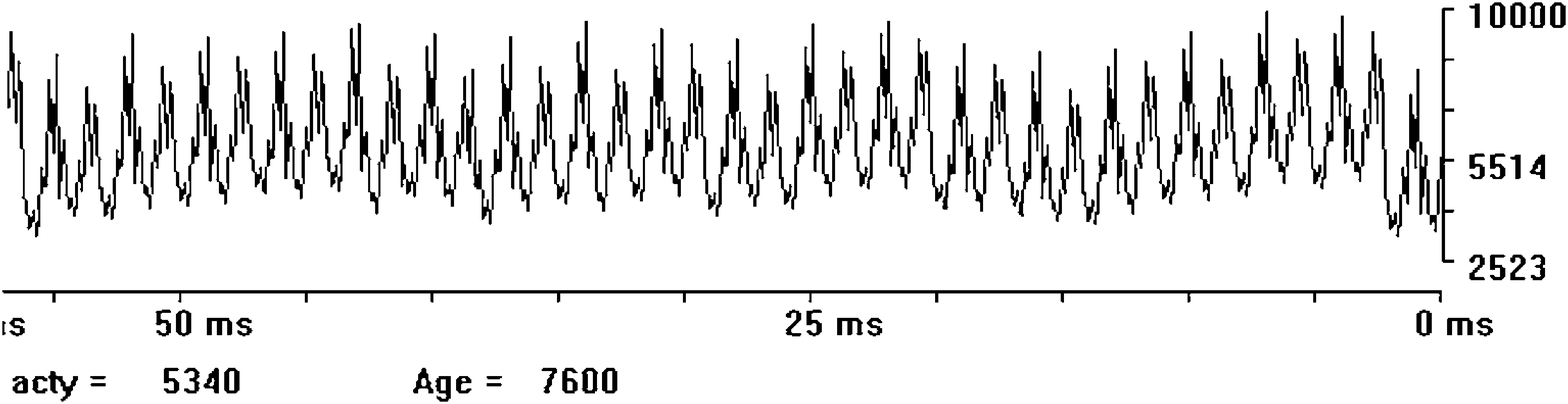}}
\put(0,75){\Large\it b}
\put(10,40){\includegraphics[width=0.69\textwidth]{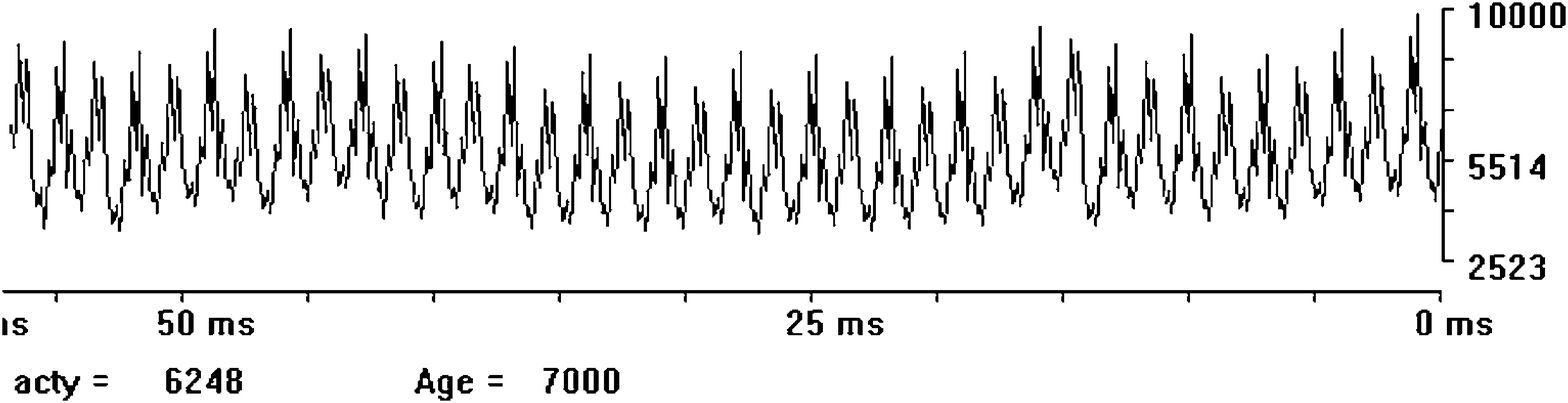}}
\put(0,35){\Large\it c}
\end{picture}
\caption{\label{acty}The network integral activity during free runs with
\protect{\it a}, $W=1$ ms,  \protect{\it b}, $W=4$ ms, and, \protect{\it c}
with $W$
interchanging from 1 to 4 ms and back every 20 ms. In the
\protect{\it c}, periods 0 ms - 20 ms, and 40 ms - 60 ms correspond to $W=4$
ms, period 20 ms - 40 ms corresponds to $W=1$ ms.
The integral activity (acty) at
any moment of time (Age) is calculated as sum of total amounts of excitation
stored in each cell and all spikes in axons at this moment of time. Age is
displayed in 0.1 ms units. Exact values of Age displayed in the panels
 correspond to point '0 ms' at the horizontal axis.  Points to the left
correspond to earlier moments. Note periodicity in \protect{\it a} with period
9 ms.} \end{figure}

\unitlength 1mm
\begin{figure}
\begin{picture}(160,130)(0,20)
\put(25,130){\includegraphics[width=0.69\textwidth]{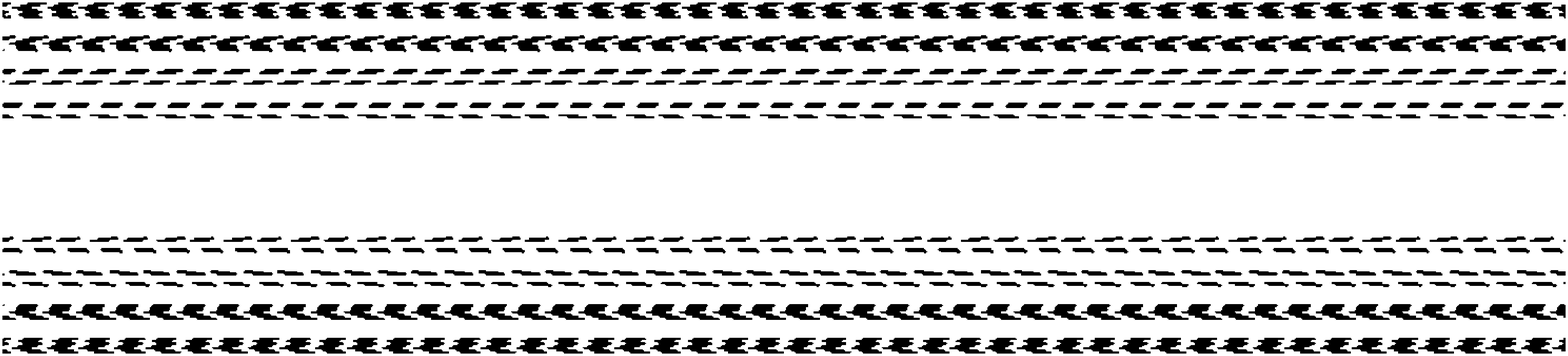}}
\put(0,147.36){\Large\it a}
\put(25,88){\includegraphics[width=0.69\textwidth]{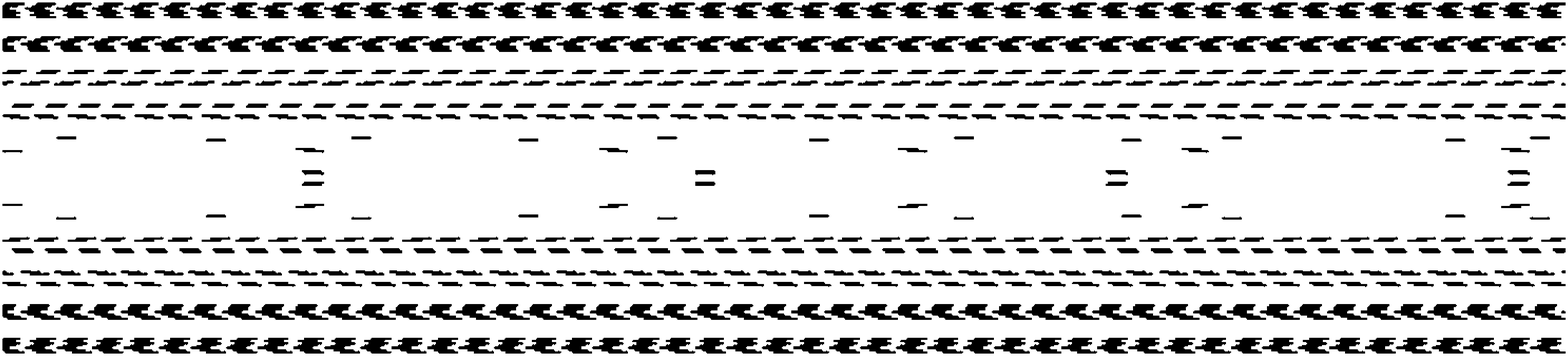}}
\put(5,130){\line(1,0){135}}
\put(17,119.65){\small 190}
\put(18.5,118.65){\vector(1,0){6.00}}
\put(0,110.36){\Large\it b}
\put(18.5,106.61){\small 94}
\put(18.5,105.61){\vector(1,0){6.00}}
\put(18.5,98.45){\small 34}
\put(18.5,97.45){\vector(1,0){6.00}}
\put(25,46){\includegraphics[width=0.69\textwidth]{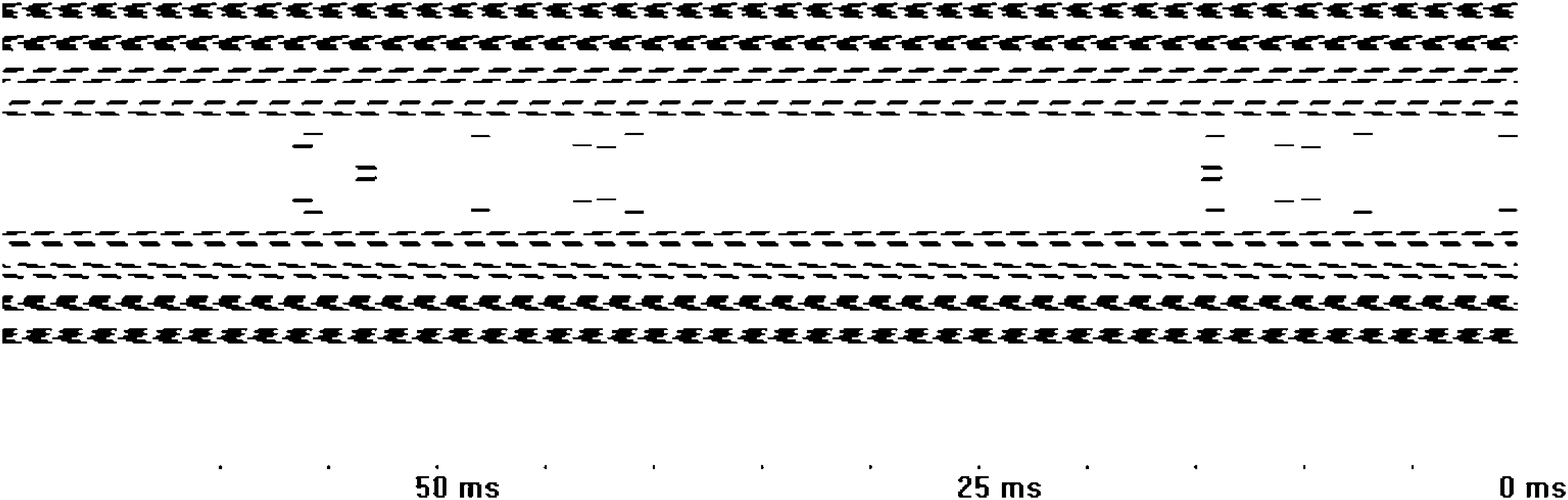}}
\put(5,88){\line(1,0){135}}
\put(0,65){\Large\it c}
\put(5,42){\line(1,0){135}}
\end{picture}
\caption{\label{refr}\protect{\it a, b, c}, the individual cells activity
during the same periods of free run as in Fig.\protect\ref{acty} panels
\protect{\it a, b, c}, respectively. The short
horizontal bars mark cells refractory period.
In the panels, cells are numbered along
the vertical axis, single pixel corresponding to a single cell.
In the \protect{\it b} panel, positions of cells \#34, \#94 and \#190 are
pointed by arrows, for clarity.  In the \protect{\it c}, periods 0 ms - 20 ms, and 40 ms -
60 ms correspond to $W=4$ ms, periods 20 ms - 40 ms, and 60 ms - 80 ms
correspond to $W=1$ ms.  } \end{figure}

After the teaching sessions both learning mechanisms were disabled. If one
allows the network to run freely after this with the locked axons staying
locked, the activity pattern in this
run resembles that used for teaching in the following sense. The pattern shown
in the Table \ref{pattern} represents a clockwise rotation of activity around
the center of the O-shaped region. In this free run activity, elements of
rotation are discriminable as well, even if exact timing of firing
deviated from the pattern used for teaching\footnote{Several teaching
protocols were tested preliminary in order to obtain a network which can
support sustained firing in a limited domains. In some of them post-teaching activity was
quiet similar to that used for teaching, but this similarity deteriorates with
time.}. Also the network integral activity (see Fig.\ref{acty}) has
oscillations similar to that in the third session of teaching.

\unitlength 1mm
\begin{figure}
\begin{picture}(160,90)(0,0)
\put(70,20.5){\includegraphics[width=0.42\textwidth]{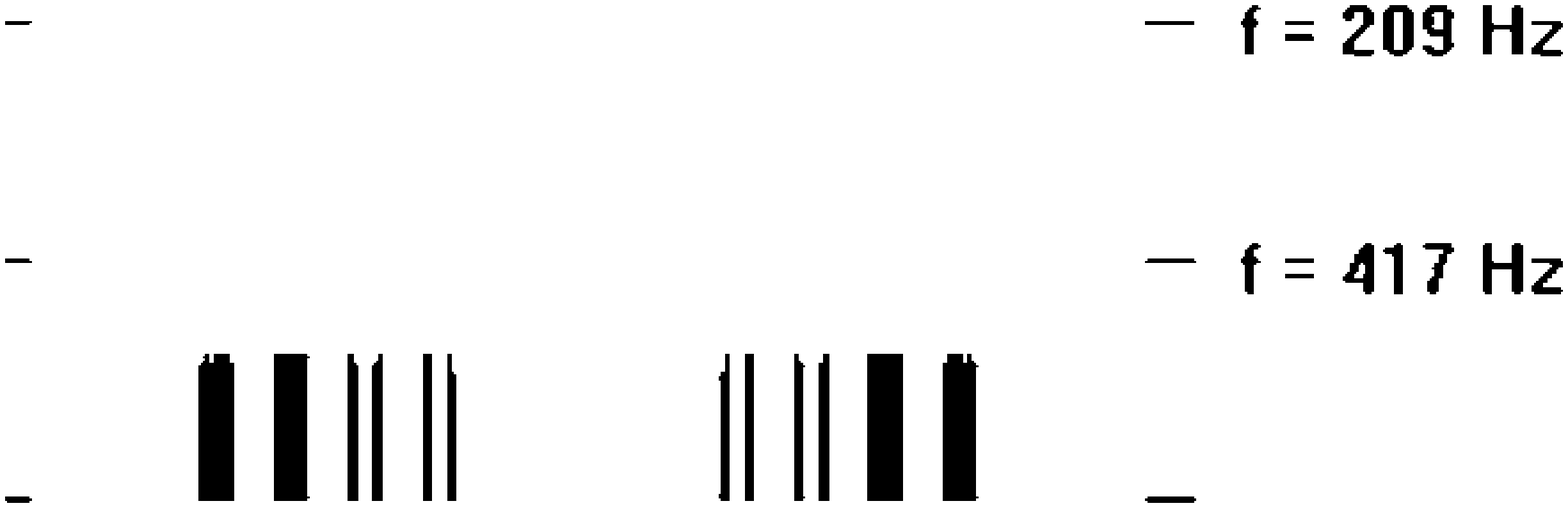}}
\put(93,7){\Large\it b}
\put(10,20.5){\includegraphics[width=0.42\textwidth]{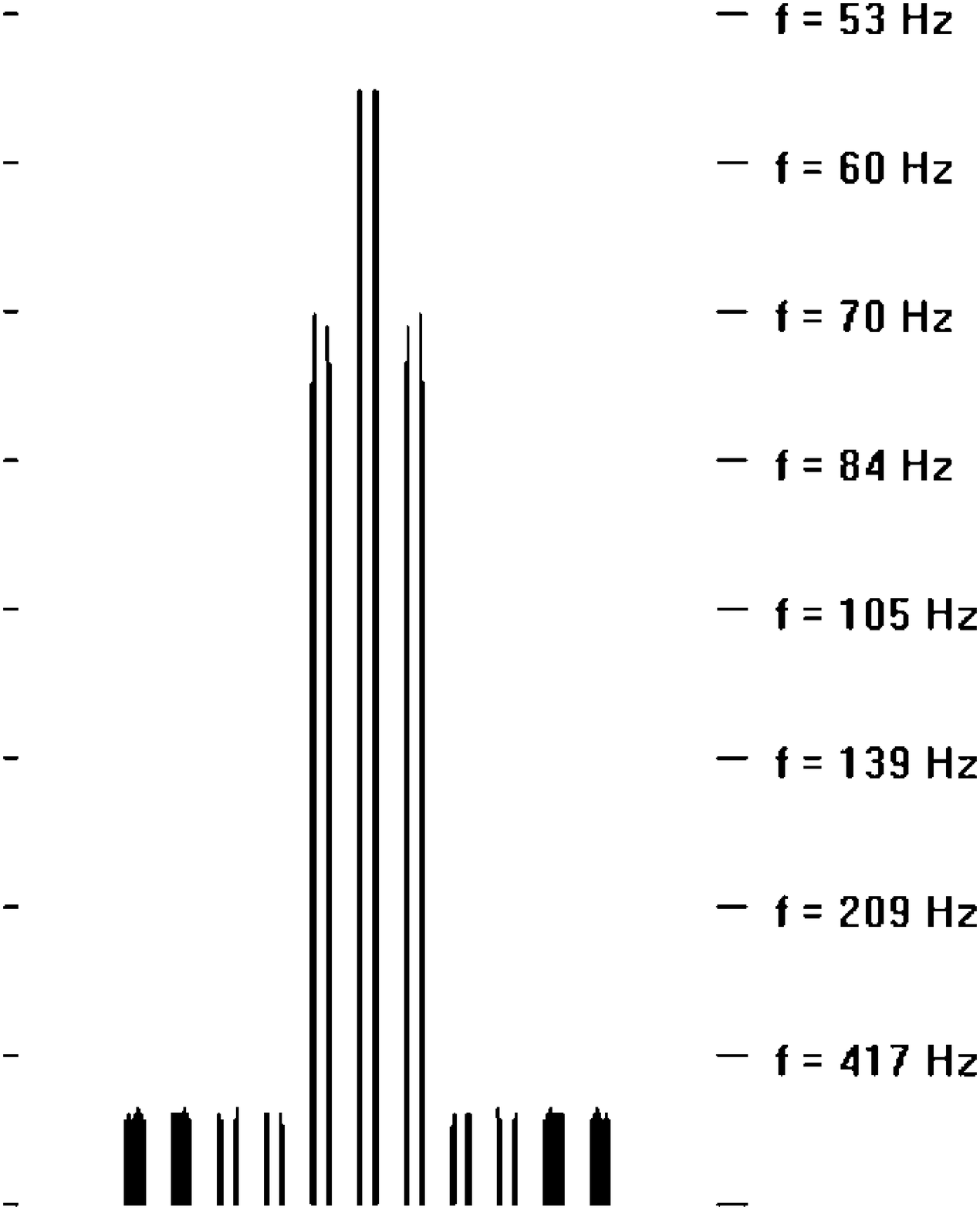}}
\put(35,7){\Large\it a}
\put(17.35,10.75){\vector(0,1){4.00}}
\put(18.35,10.75){\small 34}
\put(28.75,10.75){\vector(0,1){4.00}}
\put(30.35,10.75){\small 94}
\put(46.95,10.75){\vector(0,1){4.00}}
\put(47.55,10.75){\small 190}
\end{picture}
\caption{\label{inter}Interspike intervals for active cells in the network for
low, {\it a}, and high, {\it b} inhibition.
 The time interval between the last spike and the previous one
is shown for each cell as vertical bar with height proportional to duration
of the interval; its inversed value is specified in Hz.
 \protect{\it a, b}, correspond to $W=4$ and 1 ms, respectively.
The cells are numbered along the horizontal axis, single pixel corresponding to
 a single cell. Positions of cells \#34, \#94 and \#190 are pointed by arrows
 for clarity.
The absence of bar in some position means that
 corresponding to this position neuron was silent during period of observation.
 } \end{figure}

But, if the locked axons are made unlocked at the end of the teaching
sessions, then the dynamics brings about activity in the middle clusters.
Namely, cells in the middle clusters start firing 10 ms after beginning of the
run, and the all 12 cells are entrained into firing during next 5 ms.
If during this bound activity the inhibition is made high by
choosing $W=1$ ms, the cells in the middle clusters stop firing during first 3
ms, while activity in upper and lower clusters remains stable. The switching
between unbound and bound patterns of activity due to changed inhibition is
reproducible in the course of running dynamics, Fig.\ref{refr}.
Namely, by setting $W=1$ ms one gets the middle clusters silent (Fig.\ref{refr},
{\it a, c}), and by setting $W=4$ ms one gets firing in the middle clusters
(Fig.\ref{refr}, {\it b, c}). In the Fig.\ref{inter}, the time interval between
most recent spike and the previous one is shown for each cell as vertical bar,
its length proportional to the interval. The absence of bar at some place means
that the cell corresponding to this place was silent during time of
observation. Thus, data from Fig.\ref{inter} as well demonstrate that for high
inhibition the domain of activity is unbound (disconnected), while for low
inhibition the domain becomes bound (connected) due to activity in the middle
clusters.

\section{Conclusions and discussion}

The purpose of this paper was to discuss the binding phenomenon
and its inhibition-dependence in a simple model neural network.
The network is composed of cells of special type --- the binding neurons. Slow
inhibition in the network is expressed in terms of binding window in each cell,
as it was discussed earlier (Vidybida, 1998). The
binding neuron is chosen here as it has transparent information-processing
functionality and is simple for programming. This model is derived from
results obtained numerically for the Hodgkin-Huxley-type neuron (Vidybida,
1996). The dynamical features observed here, depend equally on the individual
cell construction and the connectivity pattern in the network as well as on
the inhibition paradigm adopted. Taking into account that binding neuron
comprises in simplified form some features of Hodgkin-Huxley neuron, one can
expect that qualitative behavior of this network should remain the same 
if individual units of the network are replaced by the H-H model, 
or leaky integrator.

By means of trials and fails, a specific teaching protocol has been found.
After applying this protocol, the network becomes able to play two patterns of
reverberating dynamics.
 In the first pattern, the neuronal activity is
confined within two spatially
disconnected domains. In the second one, the
region of activity
consists of the two above domains connected by intermediate domains into
a bound image. Raising slow inhibition switches from the bound pattern to
the disconnected one, while lowering inhibition restores the bound pattern
back, Fig.\ref{refr}. This is interpreted as switching between bound and
unbound images by means of slow inhibition.

It is difficult to offer possible physiological implications of the model
considered. The paper is aimed to discuss a possible physical mechanism, which
may underly binding. The binding itself may happen at different levels of
activity which may include different number of cells. The case of very large
number of cells may allow/require description in terms of continuous media,
with individuality of cells washed out. In any case, such dynamical feature
as entrainment into firing initially silent parts of network due to convergent
stimulation from other parts must be present in a model, because this feature
is suggested by physiological observations (see Introduction).

The choice of visually clear images of activity, as in Fig.\ref{o-net}, for the
purpose of illustrating binding can be explained as follows. Whether activity
in a neuronal population is treated as bound or unbound depends on
which device reads out from the population. If it is another neuronal
population with specific connection pattern, then two activities which are
visually similar, but have a difference discriminable by the reading
population, might be classified in the reading population as bound and unbound.
As in this work the terminal recipient of activity pattern is visual system of
a researcher, the two patterns must be visually interpretable as connected and
disconnected ones in order to be able to illustrate binding.

Mechanisms of binding may have different physical nature in different cases.
First, it could be of temporal nature (Eckhorn et al., 1988;
Engel, 1991; Vidybida, 1998), second, it could be mainly of spatial nature, as
discussed in the Introduction, third, it could be of complex spatio-temporal
nature. In this paper, only a spatial variant of binding is discussed, yet the
network constructed is suitable to study the mechanisms of all three types. For
this purpose it is necessary to modify the network by means of suitable
teaching in such a way, that it becomes able to support dynamical patterns
which are bound and unbound in temporal or spatio-temporal sense.

Finally, it should be mentioned that sometime the binding idea is not accepted
as being pertinent to information processing
(e.g., Reisenhuber and Poggio, 1999). In principle,
there are no scientific limitations for deriving all higher brain functions
immediately from activities in individual neurons. The binding concept is
useful if one wish to discuss higher functions in terms of intermediate levels
of neuronal activity. For this purpose, further elucidation of physical nature
of binding might be useful.

\noindent
{\bf References}

\hangindent=10mm\hangafter=1
Damasio, A.R., 1989. Concepts in the brain. Mind \& Language 4, 25--28.

\hangindent=10mm\hangafter=1
Damasio, H., Grabowski, T.J., Tranel, D., Hichwa, R.D., Damasio, A.R.,
1996. A neural basis for lexical retrieval. Nature 380, 499--505.

\hangindent=10mm\hangafter=1
Eckhorn, R., Bauer, R., Jordan, W., Brosch, M., Kruse, W.,
Munk, M., and Reitboeck, H.J., 1988. Coherent oscillations: a mechanism
for feature linking in the visual cortex? Biol. Cybern.
60, 121--130.

\hangindent=10mm\hangafter=1
Gerstner, W., Kempter, R., van Hemmen, J.L., Wagner, H., 1996.
A neuronal learning rule for sub-millisecond temporal coding.
Nature 383, 76--78.

\hangindent=10mm\hangafter=1
Giard, M.H., Lavikainen, J., Reinikainen, K., Perrin, F.,
Bertrand, O., Pernier, J., N\"a\"at\"anen, R., 1995. Separate
representation of stimulus frequency, intensity and duration in
auditory sensory memory: an event-related potential and dipole
model analysis. J. Cogn. Neurosci. 7, 133--143.

\hangindent=10mm\hangafter=1
Engel, A.K., Konig, P., Kreiter, A.K., Gray, C.M., and
Singer, W., 1991. Temporal coding by coherent oscillations as a
potential solution to the binding problem: physiological
evidence. In: Schuster, H.G., Singer, W. (Eds.),
Nonlinear Dynamics and Neuronal Networks. VCH Weinheim, pp. 3--25.

\hangindent=10mm\hangafter=1
Llin\'as, R., Ribary, U., Joliot, M., Wang, X.-J., 1994. Content and Context 
in Temporal Thalamocortical Binding. 
In: Buzs\'aki, G., Llin\'as, R., Singer, W.,
Berthoz, A., Christen, Y. (Eds.), Temporal Coding in the Brain.
Springer-Verlag, Berlin, pp. 251--272.

\hangindent=10mm\hangafter=1
Tranel, D., Damasio, H.,  Damasio, A.R., 1997.
A neural basis for the retrieval of conceptual knowledge. Neuropsychologia
35, 1319--1327.

\hangindent=10mm\hangafter=1
Reisenhuber, M. and Poggio, T., 1999. Are cortical models really bound by the
``binding problem"? Neuron 24, 87--93.

\hangindent=10mm\hangafter=1
Vidybida, A.K., 1996. Neuron as time coherence discriminator. Biol.
Cybern. 74, 539-544.

\hangindent=10mm\hangafter=1
Vidybida, A.K., 1998. Inhibition as binding controller at the single neuron
level. BioSystems 48, 263--267.

\hangindent=10mm\hangafter=1
von der Malsburg, C. 1999. The What and Why of Binding: 
The Modeler's Perspective. Neuron 24, 95--104.

\end{document}